\documentclass[12pt]{article}
\usepackage{amssymb,amsmath,cite,bm}
\usepackage[dvips]{graphicx,color}
\usepackage{psfrag,subfigure}
\newtheorem{proposition}{Proposition}

\newcommand{\diff}{\frac{\rm d}{{\rm d}t}}

\newcommand{\norm}[1]{\left\Vert#1\right\Vert}

\begin{document}

\title{\bf Autonomous Driving Vehicles Using Adaptive Learning Method for Data Fusion}

\author{Farhad Aghili\thanks{email: faghili@encs.concordia.ca}}

\date{}

\maketitle

\begin{abstract}
This paper presents an adaptive learning  method for data fusion in autonomous driving vehicles. 
The localization is based on the integration of Inertial Measurement Unit
(IMU) with two Real-Time Kinematic (RTK) Global Positioning System
(GPS) units in an adaptive Kalman filter (KF). The
observability analysis reveals that $i$) integration of a single GPS
with IMU does not constitute an observable system; $ii$) integration
of two GPS units with IMU results in a locally observable system
provided that the line connecting two GPS antennas is not parallel
with the vector of the measured acceleration, i.e., the sum of
inertial and gravitational accelerations. The later case makes it
possible compensate the error in the estimated orientation due
to gyro drift and its bias without needing additional instrument for
absolute orientation measurements, e.g., magnetic compass. Moreover,
in order to cope with the fact that GPS systems sometimes lose their
signal and receive inaccurate position data, the self-tuning filter
estimates the covariance matrix associated with the GPS measurement
noise. This allows the KF to incorporate GPS measurements in the
data fusion process heavily only when the information received by
GPS becomes reliably available. 
\end{abstract}

\section{Introduction}
Both position and attitude determination of a mobile robot are
necessary for navigation, guidance and steering control of a mobile
robot \cite{Aghili-Salerno-2011}. {\em Dead-reckoning} using vehicle kinematic model and
incremental measurement of wheel encoders is the common technique to
determine the position and orientation of mobile robots for indoors
applications \cite{Oryschuk-Salerno-Al-Husseini-Angeles-2009}.
However, the application of these techniques for localization of
outdoor robots is limited, particularly when the robot has to
traverse an uneven terrain or loose soils. This is because wheel
slippage and wheel imperfection cause quick accumulation of the
position and attitude errors~\cite{Kleeman-1992}. Other research
utilizes  inertial measurement unit and wheel encoders to obtain
close estimate of robot position
\cite{Vaganay-Aldon-Fourinier-1993,Fuke-Krotkov-1996,Chung-Ojeda-Borentein-2001,Dissanayake-Sukkarieh-Nebot-Durrant-Whyte-2001,Yi-Zhang-Song-Jayasuriya-2007,Lazarus-Ashokaraj-Tsourdos-2007,Yi-Wang-Zhang-2009}.
The problem with inertial systems, however, is that they require
additional information about absolute position and orientation to
overcome long-term drift~\cite{Barshan-Durrant-Whyte-1995}. Yi {\em
et al.} \cite{Yi-Wang-Zhang-2009} proposed integration of the
kinematic relationship between wheel slip and instantaneous rotation
center of skied-steered mobile robots with onboard IMU and wheel
encoder to improve motion-estimation accuracy in 2D environment.

In essence, to measure the pose of a vehicle with high bandwidth and
long-term accuracy and stability usually involves data fusion of
different sensors because there is no single sensor to satisfy both
requirements. In this respect, GPS and Inertial Measurement Unit are
considered complementary positioning systems: GPS systems provide
low update rate, but they have the advantage of long-term position
accuracy. Conversely, IMU systems provide high bandwidth position
information, while they are characterized by long term drift.
Additionally, integration of the inertial data continuously provides
pose estimation even when the GPS systems lose their signal and
receive inaccurate position data namely due to obstruction. Since no
wheel odometry is used in this localization method, one can envisage
the application of this localization method to humanoid
 and legged robots as well as aerial
vehicles \cite{Shaw-Barnes-2003}.

Nowadays, differential GPSs to centimeter-level accuracy are
commercially available making them attractive for localization,
guidance and control of outdoor mobile
robots~\cite{Bouvet-Garcia-2000,Panzieri-Pascucci-Ulivi-2002,Huang-Tan-2006,Meguro-Takiguchi-Amano-2007,Aghili-Salerno-2009,Low-Wang-2007,Low-Wang-2008,Aghili-Salerno-2011,Shair-Chandler-2008,Woo-Yoon-Cho-2009,Aghili-Salerno-2009,Asadi-Bozorg-2009,Yang-Guo-Li-2009,Aghili-Salerno-2010}.
Improving the accuracy of localization systems using RTK GPS in the
presence of GPS latency was addressed in \cite{Bouvet-Garcia-2000}.
A localization algorithm based on Kalman filtering to fuse data from
a single GPS and other several other sensors and map-based data was
presented in \cite{Panzieri-Pascucci-Ulivi-2002}. The onboard
sensory system includes wheel encodes, inertial navigation system, a
laser scanner for relative position measurements and a GPS antenna
for absolute pose measurements \cite{Aghili-2016c,Aghili-2010f}. 

An autonomous mobile robot using
GPS and photo-sensors was presented in \cite{Choi-Park-Kim-2005}.
The feasibility of a low-order vehicle positioning system
functioning under an urban environment was investigated in
\cite{Huang-Tan-2006}. This positioning system is based on
integration of Inertial Navigation System (INS) with a single GPS
unit which can provide the vehicle heading angle based on the
Doppler effect. Low {\em et al.} proposed a pose estimator using a
single RTK GPS and inertial sensors for motion estimation of a
wheeled mobile robot in 2D environment to deal with skidding and
slipping problem \cite{Low-Wang-2007,Low-Wang-2008}. However, this
sensor fusion method relies on additional instrument for absolute
orientation measurements. Similarly, a magnetic compass was
incorporated in data fusion of a MEMS-IMU/GPS integrated navigation
system proposed in \cite{Kang-hua-Mei-ping-2007} in order to make
the heading angle observable. A decentralized data fusion algorithm
is presented \cite{Asadi-Bozorg-2009} for simultaneous position
estimation of a land vehicle and building the map of the environment
by incorporating data from inertial sensor, GPS, laser scanner, the
wheel and steering encoders. The majority of the aforementioned
techniques for integration of IMU with GPS utilize a single GPS
antennas and hence they require additional instrument for absolute
orientation measurement, e.g., magnetic compass or laser scanner \cite{Aghili-Kuryllo-Okouneva-English-2010a} .
Although there are GPS devices that can provide vehicle heading
angles based on Doppler effect, the accuracy of the angle
measurement drops significantly at low speeds and it does not work
at zero speed. Data fusion from multiple sensors has been also utilized for pose estimation of vehicles in aerospace applications  \cite{Lenain-Thuilot-Cariou-2003,Aghili-Kuryllo-Okouneva-English-2010c,Asadi-Bozorg-2009,Choi-Park-Kim-2005,Aghili-Parsa-Martin-2008a,Choi-Park-Kim-2005,Aghili-Kuryllo-Okuneva-McTavish-2009,Aghili-Parsa-2007b}.

This work presents fusing data from and IMU and two RTK GPS units in
an adaptive Kalman filter to estimate the attitude, position, and
velocity  of a vehicle in three-dimensions \cite{Aghili-Salerno-2011}. Observability analysis
of GPS/IMU integration system is investigated in this paper. The
results show that the states of the system are observable provided
that at least two GPS antennas are utilized and that the line
connecting two GPS antennas is not parallel to the acceleration
measurement vector \cite{Aghili-Salerno-2016}. In other words, conventional GPS/IMU integration scheme using one GPS unit is not observable whereas the observability of the integration system using two GPS units can be ensured at the cost of adding an extra GPS unit to the integration system \cite{Aghili-2010s}. Moreover, RTK GPS devices notoriously suffer from signal robustness issue as their signal can be easily disturbed
by many factors such as satellite geometry, atmospheric condition
and shadow. To deal with the uncertain GPS noise problem, the
covariance matrix of the GPS noises is estimated in real-time so
that the KF filter incorporates GPS information heavily in the data
fusion process only when the GPS measurements become reliably
available. Tests have been conducted on the Canadian Space Agency
(CSA) {\em red rover} for assessing the performance of our pose
estimator \cite{Aghili-Salerno-2011}. This paper is organized as follow:
Section~\ref{sec:modeling} describes the observation and process
models pertaining to the positioning system consisting of two GPSs
and an IMU. Observability analysis of such a positioning system is
given in Section~\ref{sec:Observability}. In
Section~\ref{sec:Estimator}, the fusing of accelerometer, rate gyro
and GPS information in a self-tuning adaptive KF is developed.

\section{Vehicle Sensors and Modelling} \label{sec:modeling}
Fig.~\ref{fig:gps_frames} schematically illustrates a vehicle as a
rigid body to which two differential GPS-antennas and an IMU device
are attached. Coordinate frame $\{ {\cal A} \}$ is an inertial frame
while $\{ {\cal B} \}$ is a vehicle-fixed (body frame) coordinate
system. The origin of frame $\{ {\cal A} \}$ coincides with that of
the GPS base antenna, i.e., the vehicle GPS measurements are
expressed in $\{ {\cal A} \}$. Without loss of generality, we assume
that the vehicle body frame, $\{ {\cal B} \}$, coincides with the
IMU coordinate frame, i.e., the IMU measurements are expressed in
$\{ {\cal B} \}$. The orientation of $\{{\cal B}\}$ with respect to
$\{{\cal A}\}$ is represented by the unit quaternion $\bm q=[\bm
q_v^T \;\; q_o]^T$, where subscripts $v$ and $o$ denote the vector
and scalar parts of the quaternion, respectively. The rotation
matrix $\bm A$ representing the rotation of frame $\{{\cal B}\}$
with respect to frame $\{{\cal A}\}$ is related to the corresponding
quaternion $\bm q$ by
\begin{equation} \label{eq:R}
\bm A(\bm q) = (2q_o^2-1) \bm I_3 + 2q_o [\bm q_v \times] + 2 \bm
q_v \bm q_v^T,
\end{equation}
where $[\cdot \times]$ denotes the matrix form of the cross-product
and $\bm I_n$ denotes the $n\times n$ identity matrix. The
quaternion product $\otimes$ is defined as
\begin{equation*}
\bm q\otimes= q_o \bm I_4 + \bm\Omega(\bm q_v) \quad \mbox{where}
\quad \bm\Omega(\bm q_v)=\begin{bmatrix} -[\bm q_v \times] & \bm q_v
\\ -\bm q_v^T & 0 \end{bmatrix}
\end{equation*}
so that  $\bm q_1 \otimes \bm q_2$ corresponds to rotation matrix
$\bm A(\bm q_2) \bm A(\bm q_1)$.

\subsection{Suite of Sensing Systems}
This section first presents the measurement model, followed by the
process model including the close-forms of the state transition
matrix and the discrete-time process noise needed for covariance
propagation. The GPS measurements are directly included in the
measurement equations, while the IMU outputs are treated as the
time-varying inputs of the process system.

\begin{figure}[t]
\centering
\includegraphics[width=9cm]{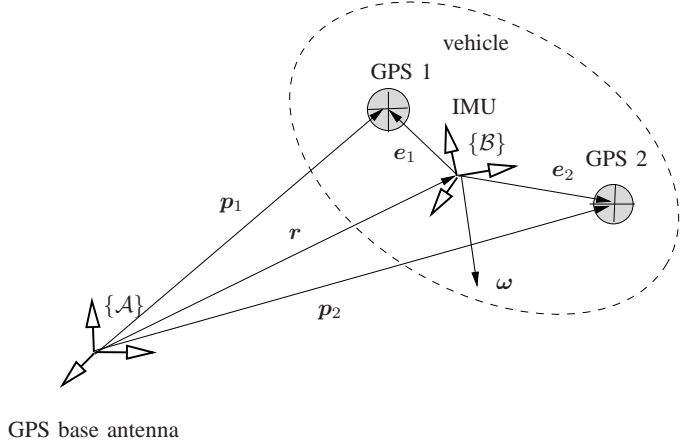}
\caption{Two GPS antennas and an IMU attached on a vehicle body.}
\label{fig:gps_frames}
\end{figure}
Assume that vector $\bm r$ represents the location of the origin of
frame $\{ {\cal B} \}$ that is expressed in coordinate frame $\{
{\cal A} \}$, and $\bm p_i$ is the output of the $i$th GPS
measurement. Apparently, from Fig.~\ref{fig:gps_frames}, we have
\begin{equation} \label{eq:pi}
\bm p_i =   \bm r + \bm A(\bm q) \bm e_i  \qquad i=1,2
\end{equation}
where constant vectors $\bm e_1$ and $\bm e_2$ are the locations of
the corresponding GPS antennas expressed in the IMU frame.

Let us define the state vector to be estimated by the EKF as $\bm x
=[\bm q_v^T \; \bm r^T \; \dot{\bm r}^T \;  \bm b^T ]^T$, where
vector $\bm b$ is  the gyro bias. Then, one can write the
observation vector as a nonlinear function of the state
\begin{equation} \notag
\bm z = \bm h(\bm x) + \bm v,
\end{equation}
where
\begin{equation} \label{eq:h}
\bm z = \begin{bmatrix} \bm p_1 \\ \bm p_2 \end{bmatrix}, \qquad \bm h(\bm x) = \begin{bmatrix} \bm r + \bm A(\bm q) \bm e_1 \\
\bm r + \bm A(\bm q) \bm e_2
\end{bmatrix},
\end{equation}
and $\bm v$ represents the GPS measurement noise, which is assumed
random walk with covariance $\bm R=E[\bm v \bm v^T ]$.

To linearize the observation vector, $\bm h(\bm x)$, one needs to
derive the sensitivity of the nonlinear observation vector with
respect to the system state vector. Consider small variation of the
quaternion from its estimation $\hat{\bm q}$ as
\begin{equation} \label{eq:delta_q}
\delta \bm q = \bm q \otimes \hat{\bm q}^{\star},
\end{equation}
where $\hat{\bm q}^{\star}$ is the inverse of quaternion $\hat{\bm
q}$, i.e., $\hat{\bm q}^{\star} \otimes \hat{\bm q} =
[0\;0\;0\;1]^T$. In other words, identity $\bm A(\bm q^{\star})
\equiv [\bm A(\bm q)]^T$ is in order for all quaternion $\bm q$.
Notice that in the following, the ``hat'' sign depicts the
estimation of a variable. Now, by virtue of $\bm A(\bm q) =\bm
A(\delta \bm q \otimes \hat{\bm q})$, one can compute the
observation vector \eqref{eq:h} in terms of the perturbation $\delta
\bm q$. Using the first order approximation of nonlinear matrix
function $\bm A(\delta \bm q)$ from expression \eqref{eq:R} by
assuming a small rotation $\delta \bm q$, i.e., $\|\delta \bm q_v\|
\ll 1$ and $\delta q_0\approx 1$, we will have
\begin{equation} \label{eq:Adeltaq}
\bm A (\delta \bm q) = \bm I_3 + 2[\delta \bm q_v \times ] +
\text{HOT}.
\end{equation}
Thus, the sensitivity matrix can be written as
\begin{equation} \label{eq:H}
\bm H = \left. \frac{\partial \bm h}{\partial \bm x} \right|_{\bm x =\hat{\bm x}}  \!\!\!
=
\begin{bmatrix}
-2 \hat{\bm A} [\bm e_1 \times]   & \bm I_3 &  \bm 0_3 &  \bm 0_3 \\
-2 \hat{\bm A} [\bm e_2 \times]    & \bm I_3 & \bm 0_3 &  \bm 0_3
\end{bmatrix},
\end{equation}
where $\hat{\bm A}=\bm A(\hat{\bm q})$.

Denoting the angular velocity of the vehicle by $\bm\omega$, the
relation between the time derivative of the quaternion and the
angular velocity can be readily expressed by
\begin{equation} \label{eq:dotq}
\dot{\bm q} = \frac{1}{2} \bm\Omega(\bm\omega) \bm q
\end{equation}
The angular rate is related to the rate gyro measurement by
\begin{equation} \notag
\bm\omega  = \bm u_g + \bm b  + \bm w_g \\
\end{equation}
where $\bm u_g$ is the gyro output, $\bm b$  is the gyro bias
vector, and $\bm w_g$ is the angular random walk noise of the gyro
rate with covariance $E[\bm w_g \bm w_g^T]=\sigma_g^2 \bm I_3$. The
time-derivative of gyro bias is traditionally modeled with random
walk model \cite{Lefferts-Markley-Shuster-1982,Pittelkau-2001}
according to
\begin{equation}\label{eq:dotb_g}
\dot {\bm b} = \bm w_{b},
\end{equation}
where $\bm w_{b}$ is the rate random walk noise with covariance
$E[\bm w_{b} \bm w_b^T]=\sigma_b^2 \bm I_3$.

A measurement of the linear acceleration of the vehicle is provided
by an accelerometer. We assume that that the deterministic error
sources of the accelerometer unit, which include scale factor and
offset, are compensated. The error can be compensated either
internally  by its signal processer or externally by a calibration
procedure \cite{Won-Golnaraghi-2010}. However, accelerometers cannot
distinguish between the acceleration of gravity and inertial
acceleration. Therefore, the accelerometer output equation is
represented by
\begin{equation}\label{eq:diff_v}
\ddot{\bm r} = \bm A(\bm q)\big(\bm u_a + \bm w_a \big) -\bm g,
\end{equation}
where  $\bm u_a$ is acceleration output, $\bm w_a$ is the
accelerometer noise assumed to be random walk noise $\bm w_a$ with
covariance $E[\bm w_a \bm w_a^T]=\sigma_a^2 \bm I_3$, and $\bm g$ is
the constant gravity vector. Therefore, in view of \eqref{eq:dotq},
\eqref{eq:dotb_g}, and \eqref{eq:diff_v} the process dynamics can be
described by
\begin{equation} \label{eq:dotx}
\dot{\bm x} = \bm f (\bm x, \bm u, \bm w )
\end{equation}
where vector $\bm u =[\bm u_g^T \; \bm u_a^T]^T$ contains the IMU
outputs, vector $\bm w=[\bm w_g^T \; \bm w_a^T \; \bm w_{b}^T
]^T$ contains the entire process noise,
\begin{equation} \notag
\bm f (\bm x, \bm u, \bm w ) = \begin{bmatrix} \frac{1}{2} \mbox{vec} \big[ \bm\Omega(\bm u_g + \bm b + \bm w_g) \bm q \big] \\
\dot{\bm r} \\ \bm A(\bm q)\big(\bm u_a + \bm w_a \big) - \bm g \\ \bm w_b
\end{bmatrix},
\end{equation}
and $\mbox{vec}(\cdot)$ returns the vector part of a quaternion.

Although the states can be propagated by solving the nonlinear
dynamics equations \eqref{eq:dotx}, the state transition matrix of
the linearized dynamics equations will be also needed to be used for
covariance propagation of the KF \cite{Aghili-Parsa-2009,Aghili-Kuryllo-Okouneva-English-2010b,Aghili-Parsa-2008b}. Adopting a linearization technique
similar to~\cite{Lefferts-Markley-Shuster-1982,Pittelkau-2001} one
can linearize \eqref{eq:dotq} about the  quaternion estimation
$\hat{\bm q}$ and $\hat{\bm\omega}= \hat{\bm u}_{g} + \hat{\bm b}$ to
obtain
\begin{subequations}
\begin{equation} \label{eq:diff_delqv}
\diff \delta \bm q_v = - \hat{\bm\omega} \times \delta \bm q_v +
\frac{1}{2} \delta \bm b + \frac{1}{2} \bm w_{g}
\end{equation}
\begin{equation}
\frac{\rm d}{{\rm d}t} \delta q_o =0
\end{equation}
\end{subequations}
Since $\delta q_o$ is not an independent variable and it has
variations of only the second order, its time-derivative can be
ignored, as suggested in~\cite{Lefferts-Markley-Shuster-1982}.

Similarly, the equation of translational motion \eqref{eq:diff_v}
can be linearized about the acceleration estimation, $\hat{\bm a}=\hat{\bm u}_a$,
and $\hat{\bm q}$ using the first-order approximation
\eqref{eq:Adeltaq} as:
\begin{align} \notag
\delta \ddot{\bm r} &= \bm A(\delta\bm{q} \otimes \hat{\bm{q}}) \big( \hat{\bm a} + \bm w_a \big) - \bm A(\hat{\bm q}) \hat{\bm a}  \\
\label{eq:diff_delr}& \approx -2 \hat{\bm A} [\hat{\bm a} \times]
\delta \bm q_v  + \hat{\bm A} \bm w_a.
\end{align}
Thus, setting \eqref{eq:dotb_g}, \eqref{eq:diff_delqv}, and
\eqref{eq:diff_delr} in the state-space form, the linearized model
of the continuous system can be derived as
\begin{subequations} \label{eq:linmodel}
\begin{equation}\label{eq:F}
\bm F = \left. \frac{\partial \bm f}{\partial \bm x} \right|_{\!\! \scriptsize \begin{array}{c}\bm x  =\hat{\bm x} \\ \bm u = \hat{\bm u} \end{array}} \!\!\!\! =
\begin{bmatrix}
- [ \hat {\bm\omega} \times] & \bm 0_3 & \bm 0_3 & \frac{1}{2} \bm I_3 \\
\bm 0_3 &  \bm 0_3  & \bm I_3 &  \bm 0_3 \\
\bm -2 \hat{\bm A}[\hat{\bm a} \times] & \bm 0_3 & \bm 0_3
& \bm 0_3 \\
\bm 0_3 &  \bm 0_3  & \bm 0_3 &  \bm 0_3
\end{bmatrix}
\end{equation}
\begin{equation} \label{eq:G}
\bm G  =   \left. \frac{\partial \bm f}{\partial \bm w}  \right|_{\bm x = \hat{\bm x}}  \!\!\!\! = \begin{bmatrix}
\frac{1}{2} \bm I_3 & \bm 0_3 & \bm 0_3
\\ \bm 0_3 & \bm 0_3  & \bm 0_3 \\\bm 0_3 & \hat{\bm A} & \bm 0_3 \\
\bm 0_3 & \bm 0_3 & \bm I_3
\end{bmatrix}
\end{equation}
\end{subequations}

\subsection{Observability Analysis} \label{sec:Observability}
The Kalman filter built around a system whose states are not
observable does not simply
work~\cite{Southall-Buxton-Marchant-1998,Aghili-2010p}. Therefore, a successful
use of Kalman filtering requires that the system be observable. A
linear time-invariant (LTI) systems is said to be {\em globally
observable} if and only if its observability matrix is full rank. If
a system is observable, the  estimation error becomes only a
function of the system noise, while the effect of the initial values
of the states on the error will asymptotically vanish. In that case,
the time-varying system \eqref{eq:H} and \eqref{eq:F} can be
replaced by a piecewise constant system for observability analysis
\cite{Goshen-Meskin-Bar-Itzhack-1992,Bryson-Sukkarieh-2008}. The
intuitive notion is that such a time-varying system can be
effectively approximated by a pieces-wise constant system without
loosing the characteristic behavior of the original system
\cite{Goshen-Meskin-Bar-Itzhack-1992,Aghili-2010d}.

The section examines the observability of the IMU/GPS integration
system for two cases: $i$) two GPS units are incorporated in the
sensor fusion; $ii$) a single GPS unit is incorporated.

\subsubsection{Adaptive Learning Data Fusion of Multiple Sensors}
The observability matrix associated with linearized system
\eqref{eq:F} together with the observation model \eqref{eq:H} is
\begin{equation} \label{eq:O}
\bm{{\cal O}} =  \begin{bmatrix} \bm H^T & (\bm H \bm F)^T & \cdots
& (\bm H \bm F^{11})^T \end{bmatrix}^T.
\end{equation}
The states of the system are assumed to be completely observable if
and only if
\begin{equation} \label{eq:rankO}
\mbox{rank}~\bm{{\cal O}}=  12
\end{equation}
which is equivalent to $\bm{{\cal O}}$ having  $12$ independent
rows. Now, let us  construct the submatrices of the observability matrix
$\bm{{\cal O}}$ from  \eqref{eq:H} and \eqref{eq:F} as
\begin{subequations} \label{eq:HFn}
\begin{align}\label{eq:HF}
\bm H \bm F &= \begin{bmatrix} 2 \hat{\bm A}[\bm e_1
\times][\hat{\bm\omega} \times]& \bm 0_3 & \bm I_3 &  -\hat{\bm A}
[\bm e_1
\times]\\
2 \hat{\bm A}[\bm e_2 \times][\hat{\bm\omega} \times]& \bm 0_3 & \bm
I_3 & -\hat{\bm A} [\bm e_2 \times]
\end{bmatrix}\\ \label{eq:HF2}
\bm H \bm F^2 &= \begin{bmatrix} -2 \hat{\bm A}([\bm e_1
\times][\hat{\bm\omega} \times]^2 +[\hat{\bm a} \times])&  \bm 0_3 &
\bm 0_3 & \hat{\bm A} [\bm
e_1 \times][\hat{\bm\omega} \times]\\
-2 \hat{\bm A}([\bm e_2 \times][\hat{\bm\omega} \times]^2+[\hat{\bm
a} \times])&  \bm 0_3 & \bm 0_3 & \hat{\bm A} [\bm e_2
\times][\hat{\bm\omega} \times]
\end{bmatrix} \\ \label{eq:HF3}
\bm H \bm F^3 & = \left[ \begin{array}{cccccc} 2 \hat{\bm A}([\bm e_1
\times][\hat{\bm\omega} \times]^3 + [\hat{\bm a} \times][\hat{\bm\omega}
\times]) & \bm 0_3 & \bm 0_3 &  -\hat{\bm A}([\bm
e_1 \times][\hat{\bm\omega} \times]^2 + [\hat{\bm a} \times]) \\
2 \hat{\bm A}([\bm e_2 \times][\hat{\bm\omega} \times]^3 + [\hat{\bm
a} \times][\hat{\bm\omega} \times]) & \bm 0_3 & \bm 0_3 &  -\hat{\bm A}([\bm e_2 \times][\hat{\bm\omega} \times]^2 + [\hat{\bm a} \times])
\end{array} \right]
\end{align}
\end{subequations}
By inspection, one can show that $\bm H \bm F^n$ with $n>0$ does not
produce any additional independent rows and therefore it is
sufficient to include only row matrices up to  $\bm H \bm F^3$ in
the observability matrix \eqref{eq:O}. As shown in the Appendix, the
observability matrix can be reduced to the following matrix by few
elementary Matrix Row Operations (MRO)
\begin{equation} \label{eq:O_MRO}
\bm{{\cal O}} \stackrel{\text{MRO}}{\longrightarrow} \bm{{\cal O}}'=
\begin{bmatrix} \bm\Pi & \bm 0_3 & \bm 0_3 & \bm 0_3\\
\times & \hat{\bm A}^T & \bm 0_3 & \bm 0_3 \\
\times & \bm 0_3 &  \hat{\bm A}^T & -[\bm e_1 \times] \\
\times &  \bm 0_3 & \bm 0_3 & \bm\Pi
\end{bmatrix},
\end{equation}
where
\begin{equation} \label{eq:Pi}
\bm\Pi=\Delta \bm e \Delta \bm e^T[\hat{\bm a} \times] + [\hat{\bm
a} \times]
\end{equation}
and vector $\Delta \bm e= \bm e_1 - \bm e_2$ is the
antenna-to-antenna baseline vector. If matrix $\bm\Pi$ is
invertible, then the reduced observability matrix $\bm{{\cal O}}'$
can be transformed into a {\em block-triangular matrix} by
pre-multiplying its fourth row by $[\bm e_1 \times]\bm\Pi^{-1}$ and
then add it to the third row. In which case, the block-triangular
matrix is full rank because all of its bock diagonal matrices are
invertible. Therefore, the full rankness of the reduced
observability matrix rests on the invertibility of the square matrix
$\bm\Pi$. In other words, if $\bm\Pi$ is invertible, then system
\eqref{eq:H}-\eqref{eq:linmodel} is observable.

\begin{proposition}
If a line connecting the two GPS antennas is not parallel with the
acceleration vector, then system \eqref{eq:H}-\eqref{eq:F} is
observable.
\end{proposition}
{\sc Proof:} In a proof by contradiction, we show that $\bm\Pi \in
\mathbb{R}^{3 \times 3}$ must be a full-rank matrix if $\hat{\bm a}
\nparallel \Delta \bm e$, i.e., vectors $\hat{\bm a}$ and $\Delta \bm e$ are not parallel. If $\bm\Pi$ is not full-rank, then there
must exist a non-zero vector $\bm\xi \neq \bm 0$ such that $\bm\Pi
\bm\xi = \bm 0_{3 \times 1}$, which can be written in this form
\begin{equation} \label{eq:xixe}
[\bm\xi \times]  \Delta \bm e - \lambda \Delta \bm e = \bm 0_{3 \times 1}\\
\end{equation}
where
\begin{equation} \label{eq:lambda}
\lambda = \bm\xi^T(\Delta \bm e \times \hat{\bm a})
\end{equation}
Notice that \eqref{eq:xixe} is the eigen equation of the
skew-symmetric matrix $[\bm\xi \times]$. The only real eigenvalue
solution of such skew-symmetric matrix is zero corresponding to
eigenvector $\bm\xi$. Therefore, substituting $\lambda=0$ and
$\Delta \bm e = \bm\xi$ in \eqref{eq:lambda} yields
\begin{equation} \label{eq:exa}
\Delta \bm e \cdot (\Delta \bm e \times \hat{\bm a}) = 0.
\end{equation}
The only possibility for nonzero vectors $\Delta \bm e$ and
$\hat{\bm a}$ to satisfy the above identity  is that the two vectors
are parallel, which is a contradiction. Thus, it is not possible for
$\bm\Pi \bm\xi= \bm 0$ to be true, meaning that matrix $\bm\Pi$ is
full rank and hence so is the observability matrix.

It is worth mentioning that the angle made by two vectors $\Delta
\bm e$ and $\hat{\bm a}$ can be calculated by
\begin{equation} \label{eq:alpha}
\theta = \cos^{-1} \frac{|\Delta \bm e \cdot \hat{\bm
a}|}{\|\Delta \bm e  \| \; \|\hat{\bm a} \|}.
\end{equation}
The above identity can be used in real-time to check if the
observability matrix is close to the ill-condition $\theta=0$.
Clearly, if the vehicle remains stationary, i.e., $\dot{\bm
r}=\ddot{\bm r} = \bm 0_{3 \times 1}$, the acceleration output
$\hat{\bm a}$ contains only the gravitational acceleration
component. In this case, the pose estimator is simply observable if
the antenna-to-antenna baseline is not parallel to the gravity
vector.

\subsubsection{Single-GPS/IMU Integration}
Now, assume that only one GPS measurement is available, say GPS~1.
Then, the sensitivity matrix matrix becomes
\begin{equation} \label{eq:H1}
\bm H =
\begin{bmatrix}
-2 \hat{\bm A} [\bm e_1 \times]   & \bm I_3 &  \bm 0_3 &  \bm 0_3
\end{bmatrix}.
\end{equation}
Consequently, the first rows of the corresponding matrices in
\eqref{eq:HFn} constitute $\bm H \bm F$ to $\bm H \bm F^3$ matrices
of the new observability matrix. By inspection, one can see that
non-zero vector
\begin{equation} \notag
\bm\eta= \begin{bmatrix} \hat{\bm a} \\ 2 \hat{\bm A}(\bm e_1 \times
\hat{\bm a}) \\ \bm 0_{3 \times 1} \\ 2 \hat{\bm\omega}\times
\hat{\bm a} \end{bmatrix}
\end{equation}
lies in the null-space of the new observability matrix $\bm{{\cal
O}}$ because
\begin{equation}
\bm{{\cal O}} \bm\eta =\bm 0_{12 \times 1}.
\end{equation}
Thais means that matrix $\bm{{\cal O}}$ is not full rank and hence
system \eqref{eq:H1}-\eqref{eq:F} is not observable. In other words,
at least two GPS antennas are required for the IMU/GPS integration
system to be observable.

\section{Adaptive Data Fusion} \label{sec:Estimator}

The equivalent discrete-time model of the linearized system
\eqref{eq:linmodel} is also required for the Kalman filter design.
The state transition matrix over time interval $t_{\Delta}$ is given
by
\begin{equation} \label{eq:expF}
\bm\Phi(t_k + t_{\Delta}, t_k)=  \bm\Phi_k = e^{\bm F(t_k)
t_{\Delta}}
\end{equation}
Then, using the sinusoidal solution of the matrix exponential of the
cross-product~\cite{Lefferts-Markley-Shuster-1982,Pittelkau-2001},
the state transition matrix $\bm\Phi_k(\tau)=\bm\Phi( t_k+\tau,
t_k)$ can be obtained by solving the matrix exponential problem
\eqref{eq:expF} as
\begin{equation*}
\bm\Phi_k(\tau) = \begin{bmatrix}
\bm\Lambda_k(\tau) & \bm 0_3 & \bm 0_3 & \frac{1}{2} \bm\Lambda'_k(\tau)\\
\bm 0_3 & \bm 0_3  &  \bm I_3 \tau & \bm 0_3 \\
-\hat{\bm A}_k [\hat{\bm a}_k \times] \bm\Lambda'_k(\tau) & \bm 0_3
& \bm 0_3 &  \bm 0_3 \\\bm 0_3 & \bm 0_3 & \bm 0_3 & \bm I_3
\end{bmatrix}.
\end{equation*}
where the submatrices of the above matrix are given by
\begin{align} \notag
\bm\Lambda_k(\tau) =& \bm I_3 - \frac{\sin\| \hat{\bm\omega}_k \|
\tau}{\| \hat{\bm\omega}_k \|}[\hat{\bm\omega}_k \times] + \frac{1 -
\cos \| \hat{\bm\omega}_k \| \tau}{\| \hat{\bm\omega}_k
\|^2}[\hat{\bm\omega}_k \times]^2
\\ \notag \bm\Lambda'_k (\tau)=&  \bm I_3 \tau +
\frac{\cos \| \hat{\bm\omega}_k \| \tau -1}{\| \hat{\bm\omega}_k \|
^2}[\hat{\bm\omega}_k \times] + \frac{\|
\hat{\bm\omega}_k \| \tau - \sin \| \hat{\bm\omega}_k \| \tau}{\|
\hat{\bm\omega}_k \|^3} [\hat{\bm\omega}_k \times]^2
\end{align}

The IMU noise constitutes the continuous process noise of the entire
system with covariance
\begin{equation} \notag
E[\bm w \; \bm w^T]=\bm\Sigma_{\rm imu}= \mbox{diag}( \sigma_g^2 \bm
I_3, \sigma_a^2 \bm I_3, \sigma_b^2 \bm I_3).
\end{equation}
The covariance matrix of the discrete-time process noise, which will
be used by the KF, can be calculated by~\cite{Jazwinski-1970}
\begin{equation} \label{eq:Q_int}
\bm Q_k = \int_{0}^{t_{\Delta}} \bm\Phi_k(\tau)\bm G \bm\Sigma_{\rm
imu} \bm G^T \bm\Phi_k^T(\tau) {\rm d} \tau.
\end{equation}
Using the first-order approximation of the matrix exponential as
$\bm\Phi_k(\tau) = e^{\bm F(t_k) \tau} \approx \bm I_{12} + \bm
F(t_k) \tau$ in \eqref{eq:Q_int} yields the covariance matrix in the
following form
\begin{equation} \notag
\bm Q_k = \begin{bmatrix} \bm Q_{k_{11}} & \times & \times & \times  \\
\bm 0_3 & \bm 0_3  & \times & \times \\
\bm Q_{k_{31}} &  \bm 0_3  & \bm Q_{k_{33}} & \times \\
\frac{1}{4} \sigma_b^2 t_{\Delta}^2  \bm I_3 & \bm 0_3 &  \bm 0_3 &
\sigma_b^2 t_{\Delta} \bm I_3
\end{bmatrix},
\end{equation}
where
\begin{align*}
\bm Q_{k_{11}} & = \big( \frac{\sigma_b^2 t_{\Delta}^3}{12}  + \frac{\sigma_g^2 t_{\Delta}}{6}\big)  \bm I_3 - \frac{\sigma_g^2 t_{\Delta}^3}{12}  [\hat{\bm\omega}_k \times]^2 \\
\bm Q_{k_{31}} & = \frac{\sigma_g^2t_{\Delta} }{4}  \bm I_3 + \frac{\sigma_g^2t_{\Delta}^2}{8} \big([\hat{\bm\omega}_k \times] -2 \hat{\bm A}_k[\hat{\bm a}_k \times]  \big)- \frac{\sigma_g^2 t_{\Delta}^3}{6} \hat{\bm A}_k [\hat{\bm a}_k \times]  [\hat{\bm\omega}_k \times] \\
\bm Q_{k_{33}} & = (\sigma_a^2  + \frac{\sigma_g^2}{4} )
t_{\Delta}\bm I_3 - \frac{\sigma_g^2 t_{\Delta}^3}{3}  \hat{\bm
A}_k [\hat{\bm a}_k \times] \hat{\bm A}_k^T 
+ \frac{\sigma_g^2 t_{\Delta}^2}{4}  \big([\hat{\bm a}_k \times]
\hat{\bm A}_k^T - \hat{\bm A}_k [\hat{\bm a}_k \times] \big).
\end{align*}

Now, one can design an extended Kalman filter based on the derived
models. Let us assume that the state vector is partitioned as $\bm
x=[\bm q_v^T \; \bm\chi^T]^T$, where $\bm\chi=[\bm r^T\; \dot{\bm r}^T \; \bm b^T]^T$ is the part of the state vector which excludes the quaternion. Then, the EKF-based observer for the
associated linearized system \eqref{eq:H}-\eqref{eq:linmodel} is
given in two steps: ($i$) estimate propagation
\begin{subequations}
\begin{align}\label{eq:state-prop}
\hat{\bm x}_{k/k-1} & = \hat{\bm x}_{k-1} +
\int_{t_k}^{t_{k}+t_{\Delta}} \bm f(\bm x, \bm u(\tau), \bm
0)\,{\text d} \tau\\ \label{eq:cov-prop} \bm P_{k/k-1}&=
\bm\Phi_{k-1} \bm P_{k-1} \bm\Phi_{k-1}^T + \bm Q_{k-1}
\end{align}
\end{subequations}
and ($ii$) estimate correction
\begin{subequations}
\begin{align} \label{eq:K_est}
\bm K_k & = \bm P_{k/{k-1}} \bm H_k^T \big(\bm H_k \bm P_{k/{k-1}} \bm H_k^T+ \bm R_k \big)^{-1} \\
\label{eq:x_est}
\Delta \hat{\bm x}_k &=
\bm K \big(\bm z_k - \bm h(\hat{\bm x}_{k/k-1}) \big) \\
\label{eq:P_est} \bm P_k &= \big( \bm I_{15} - \bm K_k \bm H_k \big)
\bm P_{k/k-1}
\end{align}
\end{subequations}
where $\hat{\bm x}_{k/k-1}$ and $\hat{\bm x}_k$ are {\em a priori} and {\em a posteriori} estimations of the state vector, and $\bm H_k =\bm H(\hat{\bm x}_{k/k-1})$. The state update follows the error-state update, $\Delta \hat{\bm x}=[\Delta \hat{\bm q}_v^T \; \Delta \hat{\bm\chi}^T ]^T$, in the {\em innovation step} of the Kalman filter \eqref{eq:x_est}. The update of ''non-quaternion part'' of the state vector, i.e. $\hat{\bm\chi}$, can be easily obtained by adding the corresponding error to a priori estimation. However, quaternion update is more complicated  because only the vectors parts of the quaternion error $\Delta \hat{\bm q}_{v_k}$ is given at every step time. Therefore, firstly, the scalar part of the quaternion variation should be computed from the corresponding vector part. Secondly, the quaternion update is computed using the quaternion multiplication rule. Consequently, the state update may proceed right after \eqref{eq:x_est} as
\begin{equation}
 \begin{split} \hat{\bm\chi}_k & = \Delta \hat{\bm\chi}_k + \hat{\bm\chi}_{k/k-1} \\
\hat{\bm q}_k & = \begin{bmatrix} \Delta \hat{\bm q}_{v_k} \\ \sqrt{1- \| \Delta \hat{\bm q}_{v_k} \|^2} \end{bmatrix} \otimes  \hat{\bm q}_{k/k-1}
\end{split}
\end{equation}

Covariance propagation in \eqref{eq:cov-prop} relies on the values
of state transition matrix of the discrete-time system, which is
linearized about the estimations of the states and the inputs at
time $t_{k-1}$, i.e., $\bm\Phi_{k-1}=\bm\Phi(\hat{\bm x}_{k-1} ,
\hat{\bm u}_{k-1})$. The sampling rate of IMU  signals are usually
higher that those of GPS signals. Therefore, the input estimation at
every step time can be obtained from averaging of IMU signals
between two consecutive GPS data acquisition in the interval $t_k <
t \leq t_k + t_{\Delta}$. That is,
\begin{equation} \notag
\hat{\bm u}_{k-1} =  \frac{1}{t_{\Delta}} \int_{t_{k-1}}^{t_{k-1} +
t_{\Delta}} \bm u(\tau) {\text d} \tau.
\end{equation}
Notice that incorporation of the decimated IMU signals in derivation
of the state transmission matrix is a good approximation to be used
only for covariance propagation. However, the position and
orientation states are propagated separately by integration of the
IMU inputs at the original sampling rate in the state propagation
step \eqref{eq:state-prop}.

\subsection{Estimation of Noise Covariance} \label{sec:noise-adaptive}

Efficient implementation of the KF requires the statistical
characteristics of the measurement and process noises. The
covariances of the IMU noises can be treated as a constant
parameter, which can be either derived from the sensor specification
or empirically tuned. However, the GPS measurement errors may vary
from one point to the next, in which case the error depends on many
factors such as satellite geometry, atmospheric condition, multipath
areas, and shadow. Therefore, since the covariance matrix associated with the GPS
noise is not known beforehand, it has to be estimated  in the filter's internal model, so that the filter is ``tuned'' as much as possible~\cite{Maybeck-1982}.

In a noise-adaptive Kalman filter, the issue is that, in addition to
the states, the covariance matrix  of the measurement noise  has to
be estimated~\cite{Maybeck-1982,Chui-Chen-1998-p113}. The fundamental assumption on which the
adaptive Kalman filter is based is that the innovation sequence can be
effectively approximated as an ergodic process inside
a sliding sampling window with length $w$. Let us define
the residual error, $\bm\varrho_k$, which is obtained from the
incoming GPS data information $\bm z_k$ and the optimal {\em a
posterior} state estimates $\hat{\bm x}_{k/k-1}$ according to
\begin{equation}
\bm\varrho_k = \bm z_k - \bm H_k\hat{\bm x}_{k/{k-1}}.
\end{equation}
The above equation can be equivalently written as
\begin{equation} \label{eq:varrho}
\bm\varrho_k = \bm H_k({\bm x}_k - \hat{\bm x}_{k/{k-1}}) + \bm v_k.
\end{equation}
Taking variance of both sides of \eqref{eq:varrho} gives
\begin{equation} \label{eq:R_hat}
\bm R_k = \bm S_k - \bm H_k \bm P_{k/{k-1}} \bm H_k^T \quad
\mbox{with} \quad \bm S_k=E[\bm\varrho_k \bm\varrho_k^T].
\end{equation}
The above equation can be used to estimate the measurement
covariance matrix $\hat{\bm R}_k$ from an an ergodic approximation
of the covariance of the zero-mean residual $\bm\varrho$ in the
sliding sampling window with finite length $w$. That is
\begin{equation} \label{eq:S_batch}
\hat{\bm S}_k  \approx \frac{1}{w}\sum_{i=k-w}^k \bm\varrho_i
\bm\varrho_i^T
\end{equation}
where $w$ is chosen empirically to give some statistical smoothing
\cite{Brown-Hwang-chapter5}. The intuitive motion in choosing a
finite window in the estimation of the innovation covariance matrix
is that very past error data has to be discounted when being used
for estimation of the current covariance.
It is known that if the innovation sequence can be assumed to be essentially time-invariant (ergodic) over the most recent $w$ steps, then using \eqref{eq:S_batch} in \eqref{eq:R_hat} yields optimal estimation of the covariance matrix $\hat{\bm R}_k$ in the sliding sampling window~\cite{Brown-Hwang-chapter5}. Then, the expression for recursive
estimation of the covariance matrix is  given by
\begin{equation} \label{eq:S_recursive}
\hat{\bm S}_{k} = \left\{ \begin{array}{ll} \frac{k-1}{k}\hat{\bm
S}_{k-1} + \frac{1}{k} \bm\varrho_k  \bm\varrho_k^T & \quad \text{if} \quad k<w\\
\hat{\bm S}_{k-1} + \frac{1}{w} \Big( \bm\varrho_k \bm\varrho_k^T -
\bm\varrho_{k-w}\bm\varrho_{k-w}^T \Big) & \quad \text{otherwise}
\end{array} \right.
\end{equation}

\begin{figure}[t]
 \centering
\includegraphics[width=12cm]{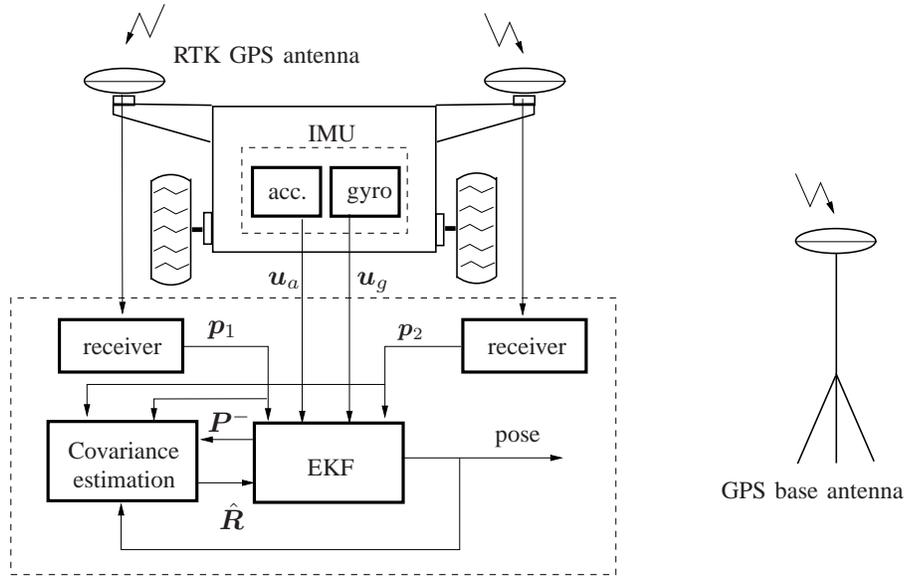} \caption{The block diagram
of the attitude determination and localization of a vehicle.}
\label{fig:gps_rover}
\end{figure}

To summarize, the adaptive estimator for driftless pose estimation
of a vehicle by fusing two RTK GPSs and IMU is schematically
illustrated in Fig.~\ref{fig:gps_rover}.

\subsection{Initialization} \label{sec:initialization}
For the first iteration of the EKF, an adequate guess of the initial
states is required.  The initial position and orientation of the
vehicle at $t=0$~s have to be carefully selected to keep the initial
error in pose estimate as small as possible based on the
information available from the measurements.

Let us form the following matrices:
\begin{equation}\label{eq:Dmatrices}
\begin{split}
\bm M &= \begin{bmatrix}  \Delta \bm p(0) & \bm g &
\Delta \bm p(0) \times \bm g  \end{bmatrix}\\
\bm N &= \begin{bmatrix}  \Delta \bm e & \hat{\bm u}_a(0) & \Delta \bm e \times
\hat{\bm u}_a(0)
\end{bmatrix}
\end{split}
\end{equation}
where $\Delta \bm p(0) \triangleq \bm p_1(0) - \bm p_2(0)$ is difference between the two GPS outputs at initial time $t=0$. In view of
\eqref{eq:pi}, one can say that vector $\Delta \bm p$ is the rotated
version of vector $\Delta \bm e$ if GPS noises are ignored.
Moreover, in static case, where $\dot{\bm r}(0) \equiv \ddot{\bm r}(0)
\equiv \bm 0$, we can say vector $\bm u_a(0)$ is the rotated version of
$\bm g$ if the accelerometer noise is ignored too. Under these
circumstances, the above two matrices are simply related by the
rotation matrix as
\begin{equation} \label{eq:ADb}
\bm M = \bm A \bm N
\end{equation}
Matrices $\bm M$ and $\bm N$ are non singular as long as $\bm g$ and
$\Delta \bm p(0)$ are not collinear, i.e., the line connecting the GPS
antennas is not parallel to the gravity vector. Then, the rotation
matrix can be obtained from matrix inversion as
\begin{equation} \label{eq:solA}
\bm A = \bm M \bm N^{-1}
\end{equation}
Solution \eqref{eq:solA} yields  a valid rotation matrix $\bm A$ so
that $\bm A^T \bm A = \bm I_3$ only if there is no error in the
column vectors of matrices  \eqref{eq:Dmatrices}. This may not be
the case in practice, however, due to the IMU and GPS noises. To
correct this problem, one may observe that all singular values of
any orthogonal matrix must be one. This means that the singular
value decomposition of the right-hand-side of \eqref{eq:solA} yields
\[ \bm M \bm N^{-1} = \bm U \bm\Sigma \bm V^T, \]
where $\bm U$ and $\bm V$ are orthogonal matrices and matrix
$\bm\Sigma=\bm I_3 + \bm\Delta_{\Sigma}$ is expected to be close to
the identity matrix, i.e., $\norm{\bm\Delta_{\Sigma}} \ll 1$.
Therefore, by ignoring small matrix $\bm\Delta_{\Sigma}$, a valid
solution for the initial rotation matrix can be found as
\begin{equation} \label{eq:AUV}
{\bm A}(0) = \bm U \bm V^T,
\end{equation}
which, then, can be used to obtain the equivalent quaternion at
$t=0$~s.

Now, with the initial guess of the rotation matrix in hand, one can
obtain the initial guess of the position from \eqref{fig:gps_frames}
as
\[{\bm r}(0) = \frac{1}{2} \big( \bm p_1(0) + \bm p_2(0) \big) - \frac{1}{2} \bm A(0)(\bm e_1 + \bm
e_2). \]

\section{Conclusions}
A adaptive learning method of data fusion in autonomous diving vehicles has been presented.  
3D attitude determination and positioning of vehicles by fusing the information from a two RTK GPS units and an IMU
in an adaptive KF has been developed. Examining the observability of
different GPS/IMU integration systems has revealed that $i$) the
single-GPS/IMU integration system is not observable, $ii$) the
dual-GPS/IMU integration system is locally observable provided that
the line connecting two GPS antennas is not parallel with the vector
of the measured acceleration. Therefore, the dual-GPS/IMU
integration method enabled the Kalman filter to compensate for the
gyro drift in the data fusion process without using additional
instrument for absolute orientation measurements, such as magnetic
compass. Moreover, since the attitude estimation method did not rely
on the heading angle measurement from the Doppler shift, it worked
at low or zero speeds. In addition to the vehicle's position and
orientation, the KF estimator was able to estimate the covariance of
the GPS measurement noises in real time. This allowed to incorporate
the GPS measurement heavily in the data fusion process only when GPS
data became reliably available. A method for adequate initialization
of the KF has been developed for fast and reliable convergence of
the estimator.

\appendix

The following matrix,
\begin{equation} \notag
\bm{{\cal O}} \stackrel{\text{MRO}}{\longrightarrow}
\begin{bmatrix} [\Delta \bm e \times ] & \bm 0_3 & \bm 0_3 & \bm 0_3\\
\bm e_1^T [\hat{\bm a} \times ] & \bm 0_{1 \times 3} & \bm 0_{1 \times 3} & \bm 0_{1 \times 3}\\
-\bm e_2^T [\hat{\bm a} \times ] & \bm 0_{1 \times 3} & \bm 0_{1 \times 3} & \bm 0_{1 \times 3}\\
\times &\hat{\bm A}^T & \bm 0_3 & \bm 0_3\\
\times & \bm 0 & \hat{\bm A}^T & -[\bm e_1 \times] \\
\times & \bm 0_3 & \bm 0_3 & [\Delta \bm e \times] \\
\times & \bm  0_{1 \times 3} & \bm 0_{1 \times 3} & \bm e_1^T[\hat{\bm a} \times] \\
\times & \bm 0_{1 \times 3} & \bm 0_{1 \times 3} & -\bm
e_2^T[\hat{\bm a} \times]
\end{bmatrix},
\end{equation}
can be constructed via the following elementary operations: The
first row of the above matrix is obtained by pre-multiplying the
first and second rows of $\bm H$ in \eqref{eq:H} by $\hat{\bm A}^T$
and then subtracting the resultant row vectors. The second and third
rows of the above matrix are obtained by pre-multiplying the first
and second rows of $\bm H \bm F^2$ in \eqref{eq:HF2} by
$-\frac{1}{2}[\hat{\bm A} \bm e_1]^T$ and by $\frac{1}{2}[\hat{\bm
A} \bm e_2]^T$, respectively. The fourth and fifth  rows are
obtained pre-multiplying the first rows of $\bm H$ in \eqref{eq:H}
and $\bm HF$ in \eqref{eq:HF} by $\hat{\bm A}^T$. The sixth row is
obtained by pre-multiplying the first and second rows of $\bm H \bm
F$ in \eqref{eq:HF} by $\hat{\bm A}^T$ and then subtracting the
resultant row vectors. Finally, the seventh and eighth rows are
obtained by pre-multiplying the first and second rows of $\bm H \bm
F^3$ in \eqref{eq:HF3} by $[\hat{\bm A} \bm e_1]^T$ and $-[\hat{\bm
A} \bm e_2]^T$, respectively. Notice that the following identity
\begin{equation} \notag
\bm e_i^T [\bm e_i \times] = \bm 0 \qquad \forall i=1,2.
\end{equation}
was used in the above derivations. Adding the second and third rows
of the above matrix and pre-multiplying the resultant row by $\Delta
\bm e$ and repeating the operation for the seventh and eighth rows
yields
\begin{equation} \label{eq:O_tall}
\stackrel{\text{MRO}}{\longrightarrow}
\begin{bmatrix} [\Delta \bm e \times ] & \bm 0_3 & \bm 0_3 & \bm 0_3\\
\Delta \bm e \Delta \bm e^T [\hat{\bm a} \times ] & \bm 0_3 & \bm 0_3 & \bm 0_3\\
\times &\hat{\bm A}^T & \bm 0_3 & \bm 0_3\\
\times & \bm 0_3 & \hat{\bm A}^T & -[\bm e_1 \times] \\
\times & \bm 0_3 & \bm 0_3 & [\Delta \bm e \times] \\
\times &  \bm 0_3 & \bm 0_3 & \Delta \bm e \Delta \bm e^T[\hat{\bm
a} \times]
\end{bmatrix},
\end{equation}
Now, one can readily show that by adding the first and second rows
of matrix \eqref{eq:O_tall} and then adding the fifth and sixth rows
of matrix \eqref{eq:O_tall}, the matrix is reduced to
\eqref{eq:O_MRO}.

\bibliographystyle{IEEEtran}

\begin{thebibliography}{10}
\bibitem{Aghili-Salerno-2011}
F.~Aghili and A.~Salerno, ``Driftless {3D} attitude determination and
  positioning of mobile robots by integration of {IMU} with two {RTK GPS}s,''
  \emph{{IEEE/ASME} Trans. on Mechatronics}, vol.~18, no.~1, pp. 21--31, Feb.
  2013.

\bibitem{Oryschuk-Salerno-Al-Husseini-Angeles-2009}
P.~Oryschuk, A.~Salerno, A.~M. Al-Husseini, and J.~Angeles, ``Experimental
  validation of an underactuated two-wheeled mobile robot,'' \emph{{IEEE/ASME}
  Trans. on Mechatronics}, vol.~14, no.~2, pp. 252--257, 2009.

\bibitem{Kleeman-1992}
L.~Kleeman, ``Optimal estimation of position and heading for mobile robots
  using ultrasonic beacons and dead-reckoning,'' in \emph{{IEEE} Int.
  Conference on Robotics \& Automation}, Nice, France, May 1992, pp.
  2582--2587.

\bibitem{Vaganay-Aldon-Fourinier-1993}
J.~Vaganay, M.~J. Aldon, and A.~Fournier, ``Mobile robot attitude estimation by
  fusion of inertial data,'' in \emph{{IEEE} Int. Conference on Robotics \&
  Automation}, Atlanta, GA, May 1993, pp. 277--282.

\bibitem{Fuke-Krotkov-1996}
Y.~Fuke and E.~Krotkov, ``Dead reckoning for a lunar rover on uneven terrain,''
  in \emph{{IEEE} Int. Conference on Robotics \& Automation}, Minneapolis,
  Minnesota, April 1996, pp. 411--416.

\bibitem{Chung-Ojeda-Borentein-2001}
H.~Chung, L.~Ojeda, and J.~Borentein, ``Accurate mobile robot dead-reckoning
  with a precision-calibrated fiber-optic gyroscope,'' \emph{{IEEE} Tran. on
  Robotics \& Automation}, vol.~17, no.~1, pp. 80--84, February 2001.

\bibitem{Dissanayake-Sukkarieh-Nebot-Durrant-Whyte-2001}
G.~Dissanayake, S.~Sukkarieh, E.~Nebot, and H.~Durrant-Whyte, ``The aiding of a
  low-cost strapdown inertial measurement unit using vehicle model constraints
  for land vehicle applications,'' \emph{{IEEE} Trans. on Robotics \&
  Automation}, vol.~17, no.~5, pp. 731--747, 2001.

\bibitem{Yi-Zhang-Song-Jayasuriya-2007}
J.~Yi, J.~Zhang, D.~Song, and S.~Jayasuriya, ``{IMU}-based localization and
  slip estimation for skid-steered mobile robots,'' in \emph{{IEEE/RSJ} Int.
  Conference on Intelligent Robots and Systems}, San Diego, CA, Oct. 29--Nov. 2
  2007, pp. 2845--2850.

\bibitem{Lazarus-Ashokaraj-Tsourdos-2007}
S.~Lazarus, I.~Ashokaraj, A.~Tsourdos, R.~Zbikowski, P.~Silson, N.~Aouf, and
  B.~A. White, ``Vehicle localization using sensors data fusion via integration
  of covariance intersection and interval analysis,'' \emph{Sensors Journal,
  IEEE}, vol.~7, no.~9, pp. 1302--1314, sep. 2007.

\bibitem{Yi-Wang-Zhang-2009}
J.~Yi, H.~Wang, J.~Zhang, D.~Song, S.~Jayasuriya, and J.~Liu, ``Kinematic
  modeling and analysis of skid-steered mobile robots with applications to
  low-cost inertial-measurement-unit-based motion estimation,'' \emph{Robotics,
  IEEE Transactions on}, vol.~25, no.~5, pp. 1087--1097, 2009.

\bibitem{Barshan-Durrant-Whyte-1995}
B.~Barshan and H.~F. Durrant-Whyte, ``Inertial navigation systems for mobile
  robots,'' \emph{{IEEE} Trans. on Robotics \& Automation}, vol.~11, no.~3, pp.
  328--342, June 1995.

\bibitem{Shaw-Barnes-2003}
A.~Shaw and D.~Barnes, ``Landmark recognition for localisation and navigation
  of aerial vehicles,'' in \emph{{IEEE/RSJ} International Conf. on Intelligent
  Robots \& Systems}, Las Vegas, Nevada, Oct. 2003, pp. 42--47.

\bibitem{Bouvet-Garcia-2000}
D.~Bouvet and G.~Garcia, ``Improving the accuracy of dynamic localization
  systems using {RTK GPS} by identifying the {GPS} latency,'' in \emph{IEEE
  Int. Conf. On Robotics and Automation}, San Francisco, CA, April 2000.

\bibitem{Panzieri-Pascucci-Ulivi-2002}
S.~Panzieri, F.~Pascucci, and G.~Ulivi, ``An outdoor navigation system using
  gps and inertial platform,'' \emph{Mechatronics, IEEE/ASME Transactions on},
  vol.~7, no.~2, pp. 134--142, jun. 2002.

\bibitem{Huang-Tan-2006}
J.~Huang and H.-S. Tan, ``A low-order {DGPS}-based vehicle positioning system
  under urban environment,'' \emph{Mechatronics, IEEE/ASME Transactions on},
  vol.~11, no.~5, pp. 567--575, oct. 2006.

\bibitem{Meguro-Takiguchi-Amano-2007}
J.~I. Meguro, J.~I. Takiguchi, Y.~Amano, and T.~Hashizume, ``{3D}
  reconstruction using multibaseline omnidirectional motion stereo based on
  {GPS}/dead-reckoning compound navigation system,'' \emph{International
  Journal of Robotics Research}, vol.~26, no.~6, pp. 625--636, 2007.

\bibitem{Aghili-Salerno-2009}
F.~Aghili and A.~Salerno, ``Attitude determination and localization of mobile
  robots using two {RTK} {GPS}s and {IMU},'' in \emph{{IEEE/RSJ} International
  Conference on Intelligent Robots \& Systems}, St. Louis, USA, October 2009,
  pp. 2045--2052.

\bibitem{Low-Wang-2007}
C.~B. Low and D.~Wang, ``Integrated estimation for wheeled mobile robot
  posture, velocities, and wheel skidding perturbations,'' in \emph{Robotics
  and Automation, 2007 IEEE International Conference on}, 2007, pp. 2355--2360.

\bibitem{Low-Wang-2008}
------, ``Gps-based tracking control for a car-like wheeled mobile robot with
  skidding and slipping,'' \emph{Mechatronics, {IEEE/ASME} Transactions on},
  vol.~13, no.~4, pp. 480--484, aug. 2008.

\bibitem{Shair-Chandler-2008}
S.~Shair, J.~H. Chandler, V.~J. Gonzalez-Villela, R.~M. Parkin, and M.~R.
  Jackson, ``The use of aerial images and gps for mobile robot waypoint
  navigation,'' \emph{Mechatronics, IEEE/ASME Transactions on}, vol.~13, no.~6,
  pp. 692--699, dec. 2008.

\bibitem{Woo-Yoon-Cho-2009}
H.~J. Woo, B.~J. Yoon, B.~G. Cho, and J.~H. Kim, ``Research into navigation
  algorithm for unmanned ground vehicle using real time kinematic
  ({RTK})-{GPS},'' in \emph{IEEE ICCAS-SICE}, Fukuoka, Japan, August 2009.

\bibitem{Asadi-Bozorg-2009}
E.~Asadi and M.~Bozorg, ``A decentralized architecture for simultaneous
  localization and mapping,'' \emph{Mechatronics, IEEE/ASME Transactions on},
  vol.~14, no.~1, pp. 64--71, feb. 2009.

\bibitem{Yang-Guo-Li-2009}
L.~Yang, Z.~Guo, Y.~Li, and C.~Li, ``Posture measurement and coordinated
  control of twin hoisting-girder transporters based on hybrid network and
  rtk-gps,'' \emph{Mechatronics, IEEE/ASME Transactions on}, vol.~14, no.~2,
  pp. 141--150, apr. 2009.

\bibitem{Aghili-Salerno-2010}
F.~Aghili and A.~Salerno, ``{3-D} localization of mobile robots and its
  observability analysis using a pair of {RTK GPS}s and an {IMU},'' in
  \emph{IEEE/ASME Int. Conf. on Advanced Intelligent Mechatronics (AIM)},
  Montreal, Canada, July 2010, pp. 303--310.

\bibitem{Aghili-2016c}
F.~Aghili and C.~Y. Su, ``Robust relative navigation by integration of icp and
  adaptive kalman filter using laser scanner and imu,'' \emph{IEEE/ASME
  Transactions on Mechatronics}, vol.~21, no.~4, pp. 2015--2026, Aug 2016.

\bibitem{Aghili-2010f}
F.~Aghili, ``Automated rendezvous \& docking {(AR\&D)} without impact using a
  reliable 3d vision system,'' in \emph{{AIAA} Guidance, Navigation and Control
  Conference}, Toronto, Canada, August 2010.

\bibitem{Choi-Park-Kim-2005}
H.-S. Choi, O.-D. Park, and H.-S. Kim, ``Autonomous mobile robot using {GPS},''
  in \emph{Int. Conference on Control \& Automation}, Budapest, Hungary, June
  2005, pp. 858--862.

\bibitem{Kang-hua-Mei-ping-2007}
T.~Kang-hua, W.~Mei-ping, and H.~Xiao-ping, ``Multiple model kalman filtering
  for {MEMS-IMU/GPS} integrated navigation,'' in \emph{Industrial Electronics
  and Applications, 2007. ICIEA 2007. 2nd {IEEE} Conference on}, may 2007, pp.
  2062--2066.

\bibitem{Aghili-Kuryllo-Okouneva-English-2010a}
F.~Aghili, M.~Kuryllo, G.~Okouneva, and C.~English, ``Fault-tolerant
  position/attitude estimation of free-floating space objects using a laser
  range sensor,'' \emph{IEEE Sensors Journal}, vol.~11, no.~1, pp. 176--185,
  Jan. 2011.

\bibitem{Lenain-Thuilot-Cariou-2003}
R.~Lenain, B.~Thuilot, C.~Cariou, and P.~Martinet, ``Adaptive control for car
  like vehicles guidance relying on {RTK GPS}: Rejection of sliding effects in
  agricultural applications,'' in \emph{IEEE Int. Conf. On Robotics and
  Automation}, Taipei, Taiwan, September 2003.

\bibitem{Aghili-Kuryllo-Okouneva-English-2010c}
F.~Aghili, M.~Kuryllo, G.~Okouneva, and C.~English, ``Fault-tolerant pose
  estimation of space objects,'' in \emph{IEEE/ASME Int. Conf. on Advanced
  Intelligent Mechatronics (AIM)}, Montreal, Canada, July 2010, pp. 947--954.

\bibitem{Aghili-Parsa-Martin-2008a}
F.~Aghili, K.~Parsa, and E.~Martin, ``Robotic docking of a free-falling space
  object with occluded visual condition,'' in \emph{9th Int. Symp. on
  Artificial Intelligence, Robotics \& Automation in Space}, Los Angeles, CA,
  Feb.~26 -- 29 2008.

\bibitem{Aghili-Kuryllo-Okuneva-McTavish-2009}
F.~Aghili, M.~Kuryllo, G.~Okuneva, and D.~McTavish, ``Robust pose estimation of
  moving objects using laser camera data for autonomous rendezvous \&
  docking,'' in \emph{{ISPRS} Worksshop Laserscanning}, Paris, France,
  September 2009, pp. 253--258.

\bibitem{Aghili-Parsa-2007b}
F.~Aghili and K.~Parsa, ``Adaptive motion estimation of a tumbling satellite
  using laser-vision data with unknown noise characteristics,'' in \emph{2007
  IEEE/RSJ International Conference on Intelligent Robots and Systems}, Oct
  2007, pp. 839--846.

\bibitem{Aghili-Salerno-2016}
F.~Aghili and A.~Salerno, \emph{Multisensor Attitude Estimation and
  Applications}, 1st~ed.\hskip 1em plus 0.5em minus 0.4em\relax CRC Press,
  2016, ch. Adaptive Data Fusion of Multiple Sensors for Vehicle Pose
  Estimation.

\bibitem{Aghili-2010s}
F.~Aghili, ``3d simultaneous localization and mapping using {IMU} and its
  observability analysis,'' \emph{Journal of Robotica}, December 2010.

\bibitem{Lefferts-Markley-Shuster-1982}
E.~J. Lefferts, F.~L. Markley, and M.~D. Shuster, ``Kalman filtering for
  spacecraft attitude estimation,'' vol.~5, no.~5, pp. 417--429, Sep.--Oct.
  1982.

\bibitem{Pittelkau-2001}
M.~E. Pittelkau, ``Kalman filtering for spacecraft system alignment
  calibration,'' vol.~24, no.~6, pp. 1187--1195, Nov. 2001.

\bibitem{Won-Golnaraghi-2010}
S.~P. Won and F.~Golnaraghi, ``A triaxial accelerometer calibration method
  using a mathematical model,'' \emph{Instrumentation and Measurement, IEEE
  Transactions on}, vol.~59, no.~8, pp. 2144--2153, 2010.

\bibitem{Aghili-Parsa-2009}
F.~Aghili and K.~Parsa, ``Motion and parameter estimation of space objects
  using laser-vision data,'' \emph{{AIAA} Journal of Guidance, Control, and
  Dynamics}, vol.~32, no.~2, pp. 538--550, March 2009.

\bibitem{Aghili-Kuryllo-Okouneva-English-2010b}
F.~Aghili, M.~Kuryllo, G.~Okouneva, and C.~English, ``Robust vision-based pose
  estimation of moving objects for automated rendezvous \& docking,'' in
  \emph{IEEE Int. Conf. on Mechatronics and Automation (ICMA)}, Xian, China,
  August 2010, pp. 305--311.

\bibitem{Aghili-Parsa-2008b}
F.~Aghili and K.~Parsa, ``An adaptive vision system for guidance of a robotic
  manipulator to capture a tumbling satellite with unknown dynamics,'' in
  \emph{{IEEE/RSJ} Int. Conf. on Intelligent Robots and Systems}, Nice, France,
  September 2008, pp. 3064--3071.

\bibitem{Southall-Buxton-Marchant-1998}
B.~B. B.~Southall and J.~Marchant, ``Controllability and observability: Tools
  for kalman filter design,'' in \emph{Proc. British Machine Vision Conference
  {(BMVC '98)}}, vol.~1, 1998, pp. 164--173.

\bibitem{Aghili-2010p}
F.~Aghili, ``Integrating {IMU} and landmark sensors for {3D SLAM} and the
  observability analysis,'' in \emph{Proc. of IEEE/RSJ International Conference
  on Intelligent Robots and Systems (IROS)}, Taipei, Taiwan, Oct. 2010, pp.
  2025--2032.

\bibitem{Goshen-Meskin-Bar-Itzhack-1992}
D.~Goshen-Meskin and I.~Y. Bar-Itzhack, ``Observability analysis of piece-wise
  constant systems. i. theory,'' \emph{Aerospace and Electronic Systems, IEEE
  Transactions on}, vol.~28, no.~4, pp. 1056 --1067, oct 1992.

\bibitem{Bryson-Sukkarieh-2008}
M.~Bryson and S.~Sukkarieh, ``Observability analysis and active control for
  airborne slam,'' \emph{Aerospace and Electronic Systems, IEEE Transactions
  on}, vol.~44, no.~1, pp. 261--280, january 2008.

\bibitem{Aghili-2010d}
F.~Aghili, ``{3D SLAM} using {IMU} and its observability analysis,'' in
  \emph{IEEE Int. Conf. on Mechatronics and Automation (ICMA)}, Xian, China,
  August 2010, pp. 377--383.

\bibitem{Jazwinski-1970}
A.~H. Jazwinski, \emph{Stochastic Processes and Filtering Theory}.\hskip 1em
  plus 0.5em minus 0.4em\relax New York: Academic International Press, 1970.

\bibitem{Maybeck-1982}
P.~S. Maybeck, \emph{Stochastic Models, Estimation, and Control (Volume
  2)}.\hskip 1em plus 0.5em minus 0.4em\relax New York: Academic Press, 1982.

\bibitem{Chui-Chen-1998-p113}
C.~K. Chui and G.~Chen, \emph{Kalman Filtering with Real-Time
  Applications}.\hskip 1em plus 0.5em minus 0.4em\relax Berlin: Springer, 1998,
  pp. 113--115.

\bibitem{Brown-Hwang-chapter5}
R.~G. Brown and P.~Y.~C. Hwang, \emph{Introduction to Random Signals and
  Applied Kalman Filtering}.\hskip 1em plus 0.5em minus 0.4em\relax John Wiley
  \& Sons, 1997, ch. The Discrete Kalman filter, State-Space Modeling, and
  Simulation, pp. 225--233.

\end{thebibliography}

\end{document}